\documentclass[12pt]{article}

\usepackage{amsmath}
\usepackage{amssymb}
\usepackage{graphicx}
\usepackage{url}
\usepackage{multirow}
\usepackage{authblk}
\usepackage{mathtools}
\usepackage{hyperref}
\usepackage[margin=1.25in]{geometry}

\newcommand\independent{\protect\mathpalette{\protect\independenT}{\perp}}
\def\independenT#1#2{\mathrel{\rlap{$#1#2$}\mkern2mu{#1#2}}}

\usepackage{tikz}
\usetikzlibrary{shapes.geometric, arrows}
\tikzstyle{box} = [minimum width=1.0cm, minimum height=1.0cm, text centered, draw=black,fill=white]
\tikzstyle{oval} = [minimum width=1.0cm, minimum height=1.0cm, text centered, draw=black,fill=white]
\tikzstyle{arrow} = [thick,->,>=stealth,line width=1]

\begin{document}

\title{Causal variance decompositions for institutional comparisons in healthcare}

\author[1]{ Bo~Chen}
\author[2]{Keith A.~Lawson}
\author[2]{Antonio~Finelli}
\author[1]{Olli~Saarela\thanks{Correspondence to: Olli Saarela, Dalla Lana School of Public Health, 155 College Street, Toronto, Ontario M5T 3M7, Canada. Email: \texttt{olli.saarela@utoronto.ca}}}

\affil[1]{Dalla Lana School of Public Health, University of Toronto}
\affil[2]{Princess Margaret Cancer Centre, University Health Network}

\maketitle

\begin{abstract}
There is increasing interest in comparing institutions delivering healthcare in terms of disease-specific quality indicators (QIs) that capture processes or outcomes showing variations in the care provided. Such comparisons can be framed in terms of causal models, where adjusting for patient case-mix is analogous to controlling for confounding, and exposure is being treated in a given hospital, for instance. Our goal here is to help identifying good QIs rather than comparing hospitals in terms of an already chosen QI, and so we focus on the presence and magnitude of overall variation in care between the hospitals rather than the pairwise differences between any two hospitals. We consider how the observed variation in care received at patient level can be decomposed into that causally explained by the hospital performance adjusting for the case-mix, the case-mix itself, and residual variation. For this purpose, we derive a three-way variance decomposition, with particular attention to its causal interpretation in terms of potential outcome variables. We propose model-based estimators for the decomposition, accommodating different link functions and either fixed or random effect models. We evaluate their performance in a simulation study and demonstrate their use in a real data application.

\end{abstract}

\section{Introduction}

\subsection{Background}

Striving for accountability and transparency in healthcare has increased the interest in quantifying and comparing the performance of institutions in terms of the quality of care that the patients receive. This is commonly done in terms of disease-specific quality indicators (QIs) that measure structural, process or outcome elements related to the care of a particular condition \cite{donabedian:1988}. For instance, in the context of surgical care for kidney cancer, examples would be (i) proportion of partial versus radical nephrectomies for early-stage patients with chronic kidney disease risk factors, (ii) proportion of minimally invasive versus open radical nephrectomies for early-stage patients, (iii) average length of stay after a radical nephrectomy or (iv) proportion of unplanned readmissions within 30 days of a radical nephrectomy \cite{wood:2013}. Process measures may be preferred for measuring quality for the reason that they are directly related to clinical decisions and thus are more actionable \cite{khuri:1998}. However, even before considering improving quality, the chosen indicator has to demonstrate existing differences in the quality of care. In doing so, it already answers a causal question, namely ``would a patient receive different care if treated in a different hospital?''. The goal of the present work is developing metrics and statistical methods that can help to identify processes or outcomes that can capture variation in the quality of care between hospitals. Following Varewyck et al. \cite{varewyck:2014}, this can be framed in a causal inference framework, where adjusting for patient case-mix factors is analogous to controlling for confounding.  However, since our focus is in choosing QIs, rather than comparing or ranking hospitals in terms of an already chosen metric, we are interested in the presence and magnitude of overall variation between the hospitals, rather than contrasts between any two hospitals. While we restrict the discussion to between hospital comparisons, without loss of generality the methods would also apply to other comparisons in the healthcare system, such as administrative subregions or individual providers (such as surgeons).

Variance decompositions have been used in the hospital profiling context to measure reliability of rankings in terms of a given quality indicator (rankability)\cite{Houwelingen1999EmpiricalBM,dishoeck:2011,Henneman:2014}, though this decomposition is constructed at aggregate (hospital) level rather than at individual patient level. This approach broadly corresponds to the random effect meta-analysis model, where the case-mix adjusted hospital-specific quality indicators $\hat \theta_z$ are assumed to follow the model $\hat \theta_z = \theta_0 + \theta_z + \xi_z$, where $z = 1, \ldots, m$ is used to index the hospital and $m$ stands for the total number of hospitals. In this model, $\theta_z \sim N(0,\tau^2)$ is a normally distributed hospital effect (`heterogeneity') around the average performance $\theta_0$, and $\xi_z \sim N(0, s_z^2)$ reflects the sampling variation under the `null' of no hospital effect ($\tau^2 = 0$). For instance, following van Houwelingen and Brand \cite{Houwelingen1999EmpiricalBM} and Racz and Sedransk \cite{racz:2010}, to obtain an indirectly standardized proportion type quality indicator, one could take $\hat \theta_z = p \frac{O_z}{E_z}$, where the overall proportion $p$ in the entire patient population is multiplied by the standardized ratio of observed to expected counts. The observed count for hospital $z$ is given by $O_z = \sum_{i=1}^n \mathbf 1_{\{Z_i = z\}} Y_i$, where $Y_i$ is the process/outcome measure of interest, and $Z_i$ indicates the hospital where patient $i \in \{1, \ldots, n\}$ was treated, with $n$ being the size of the overall patient population serving as the standard population (e.g. combination of patient populations from hospitals in a given administrative region). The expected count is given by $E_z = \sum_{i=1}^n \mathbf 1_{\{Z_i = z\}} \hat \pi_i$, where $\hat \pi_i = \textrm{expit}(\hat \alpha_0 + \hat \beta'X_i)$ is a prediction from a logistic regression model adjusted for the case-mix factors $X_i$, but not including the hospital indicators. Supposing that the standard population is large enough to ignore estimation uncertainty in the expected counts, we can take $s_z^2 = \frac{p^2}{E_z^2} \sum_{i=1}^n \mathbf 1_{\{Z_i = z\}} \hat \pi_i (1 - \hat \pi_i)$, with $\tau^2$ reflecting the between-hospital variation after adjusting for patient case-mix. The latter in turn can be estimated using the DerSimonian \& Laird method \cite{dersimonian:1987} or maximum likelihood, with the meta-analysis $I^2$ statistic indicating the proportion of variation due to between-hospital heterogeneity.

Closely related, but defined at individual level, is the concept of intra-class correlation (ICC), commonly estimated through mixed effect models \cite{Ridout:1999,Merlo:2006,Andrew2009,Yelland2011,Wu:2012,Bo:2017,Thomas2013}. Consider a generalized linear mixed effect model (GLMM)
\begin{equation} \label{gee}
E[Y_i \mid Z_i = z, X_i]= g^{-1}\left(\alpha_0 + \alpha_{z} + \beta'X_i \right),
\end{equation}
where $g()$ is the link function, and $Y_i$, $Z_i$, and $X_i$ are as before. The hospital level intercept terms are taken to follow $\alpha_{z} \sim N(0,\tau^2)$, $z = 1, \ldots, m$. For instance, with an identity link and residual variation distributed as $\varepsilon_i \sim N(0,\sigma^2)$, we have that the conditional correlation $\textrm{cor}(Y_i, Y_{i'} \mid Z_i = Z_{i'} = z, X_i, X_{i'}) = \tau^2/(\tau^2+\sigma^2)$. That is, the conditional correlation of the outcomes of two individuals treated in the same hospital is equal to the proportion of between-hospital variance of the total variance left after controlling for the patient case-mix, and can be estimated by fitting the random intercept model to obtain estimates for $\tau^2$ and $\sigma^2$. However, the connection between the ICC and the random intercept variance is more complicated with non-identity links, as the variance decomposition is different at the linear predictor and link function scale \cite{demaris:2002,Goldstein:2002}. An attempt for interpretation of the between cluster variance in terms of more familiar effect measures are the median odds ratio and median hazard ratio for binary outcomes and time-to-event outcomes, respectively \cite{austin2017}. We opt to take another route of considering the causal interpretation of the variance decomposition, as well as estimation methods that are invariant to the choice of the link function.

Furthermore, two-way variance decompositions generalize mathematically into arbitrary multi-way decompositions, though these depend on the chosen order of factorization \cite{bowsher:2012}. Aligned with our goal of identifying QIs that can capture between-hospital variation in quality without being completely driven by the case-mix, we are interested in decomposing the observed variation in care received to that explained by patient case-mix, causal between-hospital variation given the case-mix and unexplained (residual) variation.

\subsection{Objectives}
Summarizing the objectives motivated above, the structure of the paper is as follows. In Section \ref{section:notation} we recapitulate the potential outcomes (Rubin's) causal model \cite{rosenbaum:1983}, the related assumptions, and its extensions to between hospital comparisons following Varewyck et al. (2014) \cite{varewyck:2014}.  In Section \ref{section:decompositions}, we introduce the causal variance decomposition to partition observed variation in care received, and discuss its interpretation in Section \ref{section:interpretation}. In Section \ref{section:intraclass}, we investigate the connection between our causal variance decomposition and the ICC. In Section \ref{section:estimation}, we propose model-based estimation methods for the variance decompositions, and investigate their performance in a simulation study in Section \ref{section:simulation} and illustrate them in a real data application in Section \ref{section:illustration}. We conclude with a brief discussion on limitations and future directions in Section \ref{section:discussion}.

\section{Proposed measures}

\subsection{Notation and assumptions}\label{section:notation}

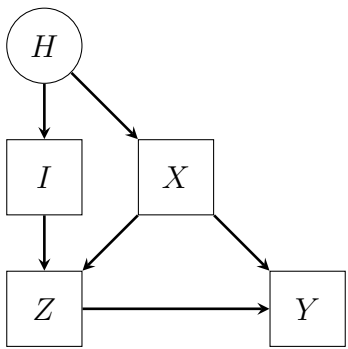
\begin{figure*}[!h]
\begin{center}
\begin{tikzpicture} [node distance=1.75cm]
\node(x) [box] {$X$};
\node(z) [box, below of=x, left of=x, xshift=0.0cm] {$Z$};
\node(y) [box, below of=x, right of=x] {$Y$};
\node(v) [box, left of=x] {$I$};
\node(h) [ellipse, oval, left of=x, above of=x, xshift=0.0cm] {$H$};
\draw [arrow] (x) -- (z);
\draw [arrow] (x) -- (y);
\draw [arrow] (z) -- (y);
\draw [arrow] (h) -- (x);
\draw [arrow] (h) -- (v);
\draw [arrow] (v) -- (z);
\end{tikzpicture}
\end{center}
\caption{Causal mechanism for hospital assignment ($Z$) and a process type indicator ($Y$). Here $X$ represents a vector of potential confounders relevant to the case-mix adjustment, while $I$ represents instrumental variables that predict the hospital assignment but are not confounders. $H$ represents latent history that can influence $I$ and $X$ but is not in itself confounder.}\label{figure:dag}
\end{figure*}

Suppressing the individual-level indices $i$ until the discussion of estimators in Section \ref{section:estimation}, let $Y \in \mathbb{R}$ represent the observed process or outcome experienced by a given patient, used to construct a quality indicator. Let $Y(z) \in \mathbb{R}$ represent the potential counterpart of this had the same patient received care in hospital $z \in \{1,...,m\}$. Let further $X=(X_{1},..,X_{p})$ be a vector of covariates relevant to the case-mix adjustment (including information on demographics, comorbidities, and disease progression), and let $Z \in \{1,...,m\}$ indicate the hospital in which the patient was actually treated. Under counterfactual consistency/stable unit treatment value assumption (SUTVA), the observed outcome is related to the potential outcomes by $Y = Y(Z)$. We assume the strong ignorability of the hospital assignment mechanism, which states that $0 < P(Z = z \mid X = x) < 1$ (positivity) and $Y(z) \independent Z \mid X$ (conditional exchangeability) at all combinations of $z$ and $x$\cite{rosenbaum:1983,Miguel:2006}. We note that as long as these conditions are satisfied, the actual mechanism which assigns patients to hospitals can be more complicated without biasing the inferences. For example, adapting from Moreno-Betancur et al.\cite{moreno:2017}, the causal relationships in the case of a process type indicator could be as illustrated in the directed acyclic graph (DAG) of Figure \ref{figure:dag}. Here the additional variables in history $H$ reflect the history (including family history) of an individual contributing to that individual living in a particular area indicated by $I$ (e.g. postal code), which in turn in part contributes to the individual being treated in a particular hospital in that area. In this DAG, $X$ are sufficient to control for confounding of the hospital effects, while $I$ are instrumental variables rather than confounders, and adjusting for them would likely lead to positivity violations.

\subsection{Three-way causal decomposition for observed variation in care received}\label{section:decompositions}

We are interested in decomposing the observed variance in the care received, $V[Y]$, to that causally explained by the quality of care differences between hospitals, by individual-level case-mix factors, and residual variation. Under the counterfactual consistency, $V[Y] = V[Y(Z)]$, for which we can write the usual two-way variance decomposition
\begin{equation} \label{eq1}
V[Y(Z)]=V_{X}[E(Y(Z) \mid X)]+E_{X}[V(Y(Z) \mid X)].
\end{equation}
For the first term, we can further write
\begin{equation}\label{eq2}
E(Y(Z) \mid X) = E_{Z \mid X}[E(Y(Z) \mid Z, X)],
\end{equation}
and for the latter term,
\begin{equation}\label{eq3}
V(Y(Z) \mid X) = V_{Z \mid X}[E(Y(Z) \mid Z, X)]+E_{Z \mid X}[V(Y(Z) \mid Z, X)].
\end{equation}
Substituting \eqref{eq2} and \eqref{eq3} into Equation \eqref{eq1}, we obtain 
\begin{equation}\label{eq4}
\begin{split}
V[Y(Z)] =& V_{X}\{E_{Z \mid X}[E(Y(Z) \mid Z,X)]\} +E_{X}\{V_{Z \mid X}[E(Y(Z) \mid Z,X)]\}\\ &+E_{Z,X}[V(Y(Z) \mid Z,X)].
\end{split}
\end{equation}
Here, due to strong ignorability, we have 
\begin{equation*}
E(Y(z) \mid Z=z,X) = E(Y(z) \mid X)  \quad\textrm{ and }\quad V(Y(z) \mid Z=z,X) = V(Y(z) \mid X).
\end{equation*}
Applying these, we get the following result:
\begin{align*}
V[Y] &= V_{X}\left\{\sum_{z}E(Y(z) \mid X) P(Z = z \mid X)\right\} \\
&\quad + E_{X} \left \{ \sum_{z}\left[E(Y(z) \mid X)  - \sum_{z'} E(Y(z') \mid X) P(Z = z' \mid X) \right]^2 P(Z = z \mid X) \right \} \\
&\quad + E_{X} \left \{ \sum_{z} V(Y(z) \mid X) P(Z = z \mid X) \right \}.
\end{align*}
We will consider the second term on the right had side as the causal quantity of interest. In this, the hospital specific performance is compared to the the expected level of care in the system for a patient with covariate values $X$, and then averaged over the covariate distribution of the standard population, similar to direct standardization. We note that this quantity is expressed in terms of potential outcomes and is a causal quantity in its own right, that is, it can be defined regardless of the causal assumptions. The causal assumptions are used to link this estimand to the observed data, working backwards to $V[Y]$. For the decomposition, we get the interpretation
\begin{align*}
\MoveEqLeft V[Y] = \textrm{total observed variance in care received}\\
&= \textrm{variance explained by the patient case-mix} \\
&\quad + \textrm{average variance causally explained by the between-hospital differences in performance} \\
&\quad\quad \textrm{conditional on case-mix} \\
&\quad + \textrm{unexplained (residual) variance}.
\end{align*}
The estimation of the components is discussed in Section \ref{section:estimation}. Briefly, the estimation of the variance decomposition can be based on modeling the components $E[Y \mid Z, X]$ and $P(Z \mid X)$ combined with the empirical distribution of $X$. Alternatively, because the decomposition can be also expressed as 
\begin{align*}
V[Y] &= V_{X}\left\{E(Y \mid X)\right\} + E_{Z,X} \left \{\left[E(Y \mid Z, X) - E(Y \mid X) \right]^2 \right \} + E_{Z,X} \left \{V(Y \mid Z, X) \right \},
\end{align*}
the estimation can also be based on modeling $E[Y \mid Z, X]$ and $E[Y \mid X]$ combined with the empirical distribution of $(Z, X)$. In Section \ref{section:estimation} we choose the former approach because it corresponds to the factorization of the likelihood, and thus enables approximate Bayesian inference. Before this, we further discuss the interpretation of the second term in the decomposition as a causal estimand.

\subsection{Causal interpretation of the decomposition}\label{section:interpretation}

To understand the causal interpretation of the variance decomposition, it is helpful to consider the special case of two hospitals, in which case the second term of the decomposition should reduce into a pairwise causal contrast. With two hospitals indexed by $z=1,2$ and denoting the propensity score $P(Z=1\mid X)=\pi(X)$, the first term of the decomposition becomes
\begin{align*}\label{2hospital1}
V_{X} \left\{ \sum_{z} E(Y(z) \mid X) P(Z = z \mid X)\right\} &= V_{X}\left\{\pi(X) E(Y(1) \mid X)+(1-\pi(X)) E(Y(2) \mid X) \right\},
\end{align*}
and the second term
\begin{equation*}\label{2hospital2}
\begin{split}
&E_{X} \left \{ \sum_{z=1}^2\left[E(Y(z) \mid X)  - \sum_{z'=1}^2 E(Y(z') \mid X) P(Z = z' \mid X) \right]^2 P(Z = z \mid X) \right \}\\
&=E_{X}\left \{ \pi(X)(1-\pi(X))\left[E(Y(1) \mid X)-E(Y(2) \mid X)\right]^2\right\} \\
&=E_{X}\left \{ V(Z \mid X)\left[E(Y(1) \mid X)-E(Y(2) \mid X)\right]^2\right\}.
\end{split}
\end{equation*}
Now, we consider three different scenarios based on the relationships between $Y$, $Z$ and $X$ in Figure \ref{figure:dag}.
\begin{description}
\item[Scenario 1.] In the absence of the arrow $X\rightarrow Y$, which implies $Y \independent X \mid Z$ and $ E(Y(z) \mid X) = E(Y(z))$, the first term is equal to $V_{X}(\pi(X))\left[E(Y(1))-E(Y(2))\right]^2 $. This implies that this term takes the variance contribution of the indirect (mediated) pathway $X \rightarrow Z \rightarrow Y$, which is present only if $Z$ has causal effect on $Y$. The second term is equal to $E_{X}[V(Z \mid X)] [E(Y(1))-E(Y(2))]^2$. The difference in interpretation between these two terms is that the former represents the variance due to different kinds of patients being treated by different hospitals, while the latter represents the variation in care similar patients receive. If all hospitals treat similar patient populations, the former term disappears, while the latter is maximized. In the opposite case where the hospitals specialize in treating different patient populations, the positivity assumption is violated and the process/outcome cannot reveal performance differences between the hospitals. Thus, we focus on the second term as a measure of goodness of a proposed process/outcome as a QI.

\item[Scenario 2.] In the absence of the arrows $X \rightarrow Z$ and $I \rightarrow Z$, which implies $Z \independent X$ and $\pi(X)=\pi$ (randomized assignment), the first term would be equal to $V_{X}\left\{\pi [E(Y(1) \mid X)-E(Y(2) \mid X)]+E(Y(2) \mid X)\right\}$. The two additive components inside the variance reflect the effect modification by $X$ and the predictive variance due to $X$. In the absence of effect modification/additive interaction, that is, $E(Y(2) \mid X)=E(Y(1) \mid X) + \alpha$, it follows that $\pi [E(Y(1) \mid X)-E(Y(2) \mid X)] = \pi\alpha$ is constant and only the predictive variance remains. The second term is equal to $V(Z)E_{X}\left \{\left[E(Y(1) \mid X)-E(Y(2) \mid X)\right]^2\right\}$. In the absence of effect modification, this becomes $\pi(1-\pi)\alpha^2$, capturing the causal effect. Compared to the usual direct standardization formula $E(Y(1))-E(Y(2)) = E_X[E(Y(1) \mid X)-E(Y(2) \mid X)]$, the between-hospital variance is similarly averaged over $X$, but squared so that any contrast at the same level of $X$ contributes to the variance rather than canceling out. The reason effect modification contributes to both case-mix and between-hospital variance components is that it both increases the variance of the expected care level (non-causal) and the average squared difference from the expected care level (causal). When evaluating proposed processes/outcomes, we would again want the second term to be comparatively large compared to the first one, because even though the eventual QI will be case-mix adjusted, large case-mix variance indicates that the measure is less intervenable, as the variation in care largely reflects treatment choices necessitated by disease progression, comorbidities etc., and makes the indicator sensitive to the method of case-mix adjustment.

\item[Scenario 3.] In the absence of the arrow $Z \rightarrow Y$, which implies $Y \independent Z \mid X$ and $E(Y(1) \mid X) = E(Y(2) \mid X)$, the first term becomes  $V_{X}[E(Y(2) \mid X)]$, reflecting the predictive variance due to $X$, while the second term is equal to zero as it should in the absence of causal effect of $Z$ on $Y$.
\end{description}

\subsection{Connection to intra-class correlation}\label{section:intraclass}

We note that the previously derived variances were weighted by the quantities $P(Z = z \mid X)$ reflecting the hospital patient volumes; this is natural since we are decomposing the observed variance in the entire combined patient population, and larger centers carry more weight in this. However, to expand the proposed framework, we can consider variance decompositions under hypothetical assignment mechanisms that can differ from the actual one. For example, referring to the interpretation of Scenario 1 of Section \ref{section:interpretation}, we can consider the variance decomposition under a hypothetical `randomized' assignment mechanism where every hospital treats similar kind of patient population. This also allows us to derive a connection between the causal between-hospital variance in the previous section, and the more familiar intraclass correlation. Let $A$ represent a random draw from such a hypothetical assignment mechanism, with the mechanism chosen so that $A \independent Z \mid X$, and strong ignorability $0 < P(A = a \mid X = x) < 1$ and $Y(a) \independent A \mid X$ apply. This lets us consider random draws of potential outcomes $Y(A)$, similar to the notation used by VanderWeele et al.\cite{VanderWeele:2014}. Now, because
\begin{equation*}
E(Y(a) \mid A=a,X) = E(Y(a) \mid X) = E(Y(a) \mid Z=a,X),  
\end{equation*}
we can write the variance decomposition
\begin{align*}
V[Y(A)] 
&= V_{X}\left\{\sum_{a}E(Y(a) \mid X) P(A = a \mid X)\right\} \\
&\quad + E_{X} \left \{ \sum_{a}\left[E(Y(a) \mid X)  - \sum_{a'} E(Y_i(a') \mid X) P(A = a' \mid X) \right]^2 P(A = a \mid X) \right \} \\
&\quad + E_{X} \left \{ \sum_{a} V(Y(a) \mid X) P(A = a \mid X) \right \}.
\end{align*}
If we choose here $P(A = a \mid X) =  P(Z = a \mid X) \;\forall\; a \in \{1, \ldots, m\}$, the result is equivalent to before. However, if we choose to weight the hospitals equally as $P(A = a \mid X) =  1/m$, and assume a linear model 
\begin{align*}
E(Y(a) \mid Z=a,X) = E(Y(a) \mid X) = \alpha_0 + \alpha_a + \beta'X,
\end{align*}
we can express the second term as
\begin{align}\label{equation:icc}
\MoveEqLeft E_{X} \left \{ \sum_{a}\left[E(Y(a) \mid X)  - \sum_{a'} E(Y(a') \mid X) P(A = a' \mid X) \right]^2 P(A = a \mid X) \right \} \nonumber \\
&= E_{X} \left \{ \frac{1}{m} \sum_{a}\left[E(Y(a) \mid X)  - \frac{1}{m}\sum_{a'} E(Y(a') \mid X) \right]^2 \right \} \nonumber \\
&= \frac{1}{m} \sum_{a}\left[\alpha_a  - \frac{1}{m}\sum_{a'} \alpha_{a'} \right]^2.
\end{align}
If we take the hospital effects to represent a `sample' from a distribution with variance $V[\alpha_a] = \tau^2$, \eqref{equation:icc} converges to $\tau^2$ when $m \rightarrow \infty$. In reality, since our discussion is in the administrative data context, we observe all the hospitals in the administrative region of interest and don't take them to be a sample from a superpopulation of hospitals. Nevertheless, the result serves to establish a connection to the ICC: because
 \begin{align*}
V(Y(A) \mid X) &= V_{A \mid X}[E(Y(A) \mid A, X)]+E_{A \mid X}[V(Y(A) \mid A,X)] \\
&\rightarrow  \tau^2 + \sigma^2 \quad\textrm{when}\quad m \rightarrow \infty,
\end{align*}
under a linear model the causal interpretation of $ICC=\frac{\tau^2}{\tau^2 + \sigma^2}$ is the proportion of case-mix conditional variance that is causally explained by the between-hospital variation in practices. However, this interpretation is specific to the linear model with equal hospital weights, which is why we proceed with the decomposition in Section \ref{section:decompositions} for the observed variance; this is general and can be estimated in the same way irrespective of the outcome model formulation (that is, irrespective of the choice of the link function, or whether the hospital effects are modeled as fixed or random effects).

\section{Estimators}\label{section:estimation}

\subsection{Point estimators}

We present model-based direct standardization type estimators for the variance components. To model the outcomes, we specify a generalized linear model
\begin{equation*}
E[Y(z) \mid X; \theta]=E[Y \mid Z=z, X; \theta]= g^{-1}\left(\alpha_0+\alpha_{z} + \beta'X \right),
\end{equation*}
where $\theta = (\alpha_0, \alpha_1, \ldots, \alpha_m, \beta)$. The hospital level intercept terms $\alpha_{z}, z\in\{1,...,m\}$ are either fixed, with $\alpha_1 = 0$, or follow $\alpha_{z} \sim N(0,\tau^2)$ (random intercept model). As noted in the previous section, the random effects are adopted as a means to apply shrinkage to the hospital effects, rather than a data generating mechanism. With random effects, the fitted values are obtained using empirical Bayes prediction for the random intercepts. We note that in the model we can also introduce hospital-case-mix interactions, as discussed by Varewyck et al.\cite{Varewyck:2016}, but omit these here for notational simplicity. We estimate the parametrized variance explained by the patient case-mix as
\begin{align*}
\omega_1(\theta, \eta) &= V_{X}\left\{\sum_{z}E(Y(z) \mid X; \theta) P(Z = z \mid X; \eta)\right\} \\
&\approx  \frac{1}{n-1} \sum_{i=1}^n \left\{\sum_{z}\mu_i(z; \hat \theta) P(Z_i = z \mid x_i; \hat \eta) - \frac{1}{n} \sum_{i'=1}^n \sum_{z} \mu_{i'}(z, \hat \theta) P(Z_{i'} = z \mid x_{i'}; \hat \eta) \right\}^2,
\end{align*}
where $\mu_i(z; \theta) = E[Y_{i} \mid Z_i=z, x_i; \theta]$ and where the terms $P(Z = z \mid X; \eta)$ can be estimated by fitting a multinomial logistic regression model 
\begin{equation}\label{equation:multinomial}
P(Z=z \mid X; \eta) =
\begin{cases}
\frac{1}{1+\sum_{z'=2}^m \exp(\gamma_{z'}+\phi_{z'}'X)} & z=1\\
\frac{ \exp(\gamma_{z}+\phi_z'X)}{1+\sum_{z'=2}^m \exp(\gamma_{z'}+\phi_{z'}'X)}& z\neq1,
\end{cases}
\end{equation}
where $\eta = (\gamma_2, \ldots, \gamma_m, \phi_2, \ldots, \phi_m)$. Similarly, the parametrized average variance explained by the hospital performance conditioning on patient case-mix can be estimated by
\begin{align*}
\omega_2(\theta, \eta) &= E_{X} \left \{ \sum_{z}\left[E(Y(z) \mid X; \theta)  - \sum_{z'} E(Y(z') \mid X; \theta) P(Z = z' \mid X; \eta) \right]^2 P(Z = z \mid X; \eta) \right \} \\
&\approx \frac{1}{n} \sum_{i=1}^n \left\{\sum_{z}\mu_i(z; \hat \theta)^2 P(Z_i=z \mid x_i; \hat \eta)-\Bigg[\sum_{z}\mu_i(z; \hat \theta) P(Z_i = z \mid x_i; \hat \eta)\Bigg]^2\right\}.
\end{align*}
The residual variance can then be estimated by subtracting the two estimated components from the empirical total variance, or using a distributional assumption such as 
$V(Y_i \mid Z_i = z, x_i; \theta) = \mu_i(z; \theta)[1 - \mu_i(z; \theta)]$ in the case of binary indicator, and taking
\begin{align*}
\omega_3(\theta, \eta) &= E_{X} \left \{ \sum_{z} V(Y(z) \mid X; \theta) P(Z = z \mid X; \eta) \right \} \\
&\approx
\frac{1}{n}\sum_{i=1}^n \left \{ \sum_{z} \mu_i(z; \hat\theta)[1-\mu_i(z; \hat\theta)] P(Z = z \mid x_i; \hat\eta) \right \}.
\end{align*}
\subsection{Variance estimators}

To quantify the uncertainty in the variance components, we note that conditional on the empirical covariate distribution in the standard population the estimates are entirely model-based. Thus, we can produce approximate samples from the joint posterior distribution 
\begin{align*}
p(\theta, \eta \mid \mathbf{X}, \mathbf{Y}, \mathbf{Z}) &= \frac{p(\mathbf{Y}, \mathbf{Z} \mid \theta, \eta, \mathbf{X}) p(\theta \mid \mathbf{X}) p(\eta \mid \mathbf{X})}{p(\mathbf{Y}, \mathbf{Z} \mid \mathbf{X})} \\
&=  \frac{p(\mathbf{Y} \mid \theta, \mathbf{X}, \mathbf{Z}) p(\theta \mid \mathbf{X})}{p(\mathbf{Y} \mid \mathbf{X}, \mathbf{Z})} 
\frac{p(\mathbf{Z} \mid \eta, \mathbf{X}) p(\eta \mid \mathbf{X})}{p(\mathbf{Z} \mid \mathbf{X})} \\
&= p(\theta \mid \mathbf{X}, \mathbf{Y}, \mathbf{Z}) p(\eta \mid \mathbf{X}, \mathbf{Z}),
\end{align*}
by sampling $\theta$ and $\eta$ separately from their respective posteriors, and recalculate the variance components $\omega_1(\theta, \eta)$ and $\omega_2(\theta, \eta)$ at each combination of $\theta$ and $\eta$ to quantify uncertainty in them. Under flat priors the sampling distribution variances of the parameter estimates approximate the respective posterior variances; in approximate sampling of $\theta$ we used parametric bootstrap to reflect the uncertainty in the random effect estimates; for example for a binary indicator we can repeatedly sample the outcomes from $Y_i \mid (z_i, x_i; \hat \theta) \sim \textrm{Bernoulli}(\mu_i(z_i; \hat\theta))$, $i = 1, \ldots, n$, and refit the mixed model to produce a sample of the model-based fitted values $\mu_i(z; \theta)$. Compared to non-parametric bootstrap this avoids some hospital categories being absent in some of the replicates. For sampling of $\eta$ we used the normal approximation $\eta \mid (\mathbf{X}, \mathbf{Z}) \sim N(\hat\eta, V(\hat\eta))$ where $\hat\eta$ is the maximum likelihood estimate of $\eta$ and $V(\hat\eta)$ its asymptotic variance-covariance matrix from the multinomial logistic model fit.

\section{Simulation study}\label{section:simulation}

\subsection{Generating mechanism}

We demonstrate the performance of the estimators by simulating data from a generating mechanism simplified (omitting $H$ and $I$) from that in Figure \ref{figure:dag}, varying the total number of patients $n$ and number of hospitals $m$. Our objectives in the simulation are (a) to demonstrate that the proposed estimators work both for linear models for continuous outcomes and logistic models for dichotomous outcomes, (b) to demonstrate that the estimators work both with fixed and random effect models, and (c) to obtain evidence of their asymptotic behaviour under increasing $n$ and $m$. First, we generated two case-mix factors from $X_{1}\sim N(0, 1)$ and $X_{2} \sim \textrm{Bernoulli}(0.5)$, given which the hospital assignment $Z$ was generated from multinomial logistic model specified as in \eqref{equation:multinomial}, drawing the intercepts from $\gamma_z \sim N(0, 0.25)$ and the regression coefficients from $\phi_{z1} \sim N(0, 0.5)$ and $\phi_{z2} \sim N(0, 0.5)$, $z = 2, \ldots, m$. In our generating mechanism, we did not consider the hospitals themselves to be a sample from a superpopulation of hospitals, and thus we fixed the hospital-specific parameters across the replications (i.e. drawing these only once from the above distributions), while resampling the patients in each replication. This corresponds to conventional analysis of administrative data, where all the hospitals operating in a given region are compared, but the observed patient population is taken to represent a sample from an unobserved``long run''. Outcomes were generated from the mean structure
\begin{equation}\label{eq11}
g(E[Y(z) \mid X; \theta]) = \alpha_{z} + X_{1} + 2 X_{2},
\end{equation}
with the hospital effects drawn from $\alpha_z \sim N(0, 2^2)$ (again fixing these parameter values across the replications). Continuous potential outcomes were generated by taking $Y(z) = E[Y(z) \mid X; \theta] + \varepsilon$, where $\varepsilon \sim \textrm{Logistic(0,1)}$, with the observed outcome given by $Y(Z)$. Binary outcomes were generated by dichotomizing at zero as $\tilde Y(z) = \mathbf 1_{\{Y(z) \ge 0\}}$. To calculate the estimators, correctly specified outcome and assignment models were fitted to each simulated dataset, with both fixed and random effect specifications used for the former, whereas the true decomposition was calculated using the known values fixed in the data generating mechanism.

\subsection{Results}

\subsubsection{Continuous Outcomes}

\begin{figure}[!h]
\centering
\includegraphics[width=130mm]{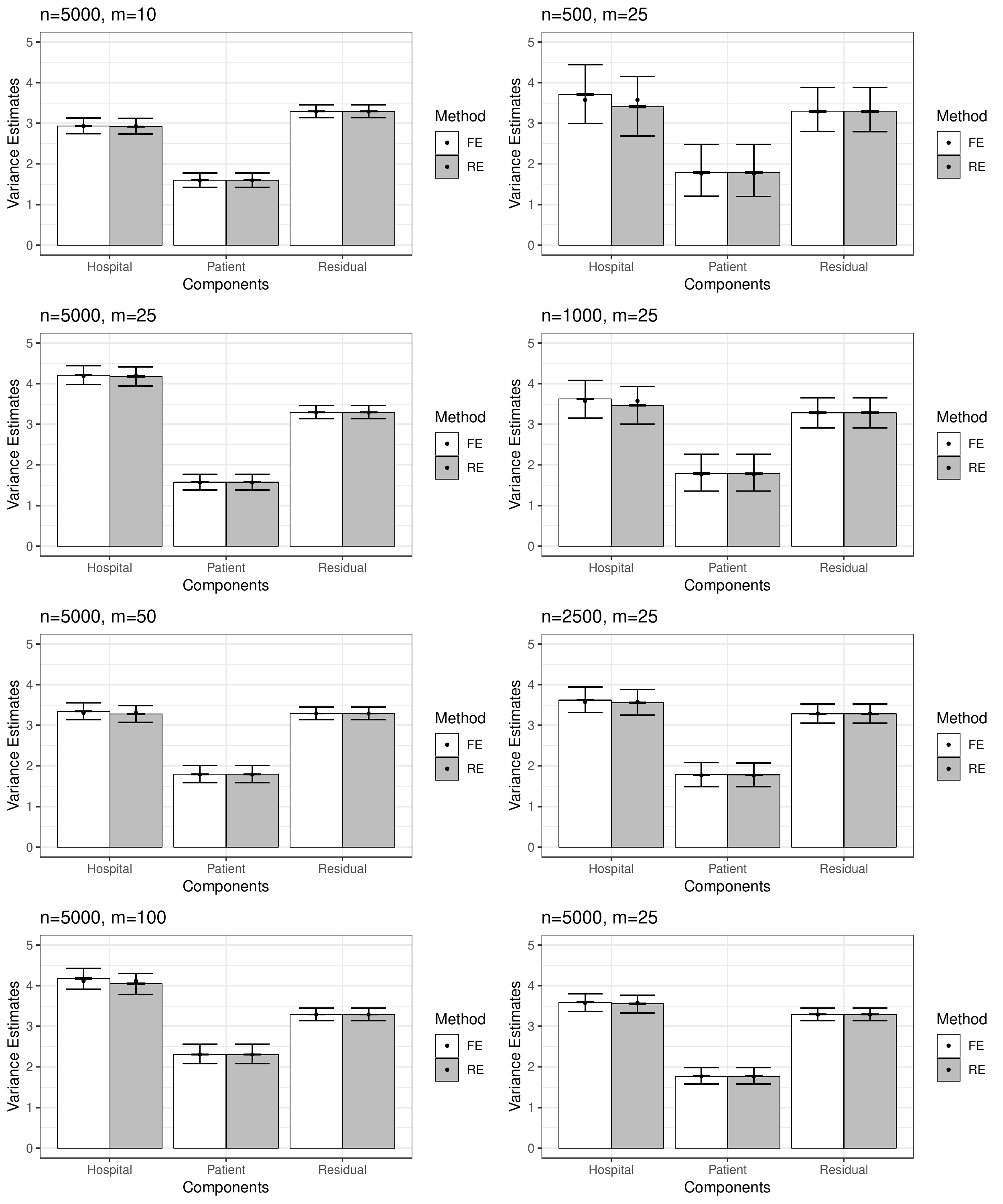}
\caption{Simulated sampling distribution means for the three variance components under both fixed and random effect models for the continuous outcomes under different combinations of $n$ (total number of patients) and $m$ (total number of hospitals), based on 1000 replications. The black dots indicate the true variances. Also shown are the 95\% quantile interval for the sampling distribution, and the 95\% confidence interval for the mean, reflecting the Monte Carlo error in the estimated mean of the sampling distribution. FE and RE stand for fixed-effect and random-effect models.}
\label{Plot1_con}
\end{figure}

\begin{figure}[!h]
\centering
\includegraphics[width=130mm]{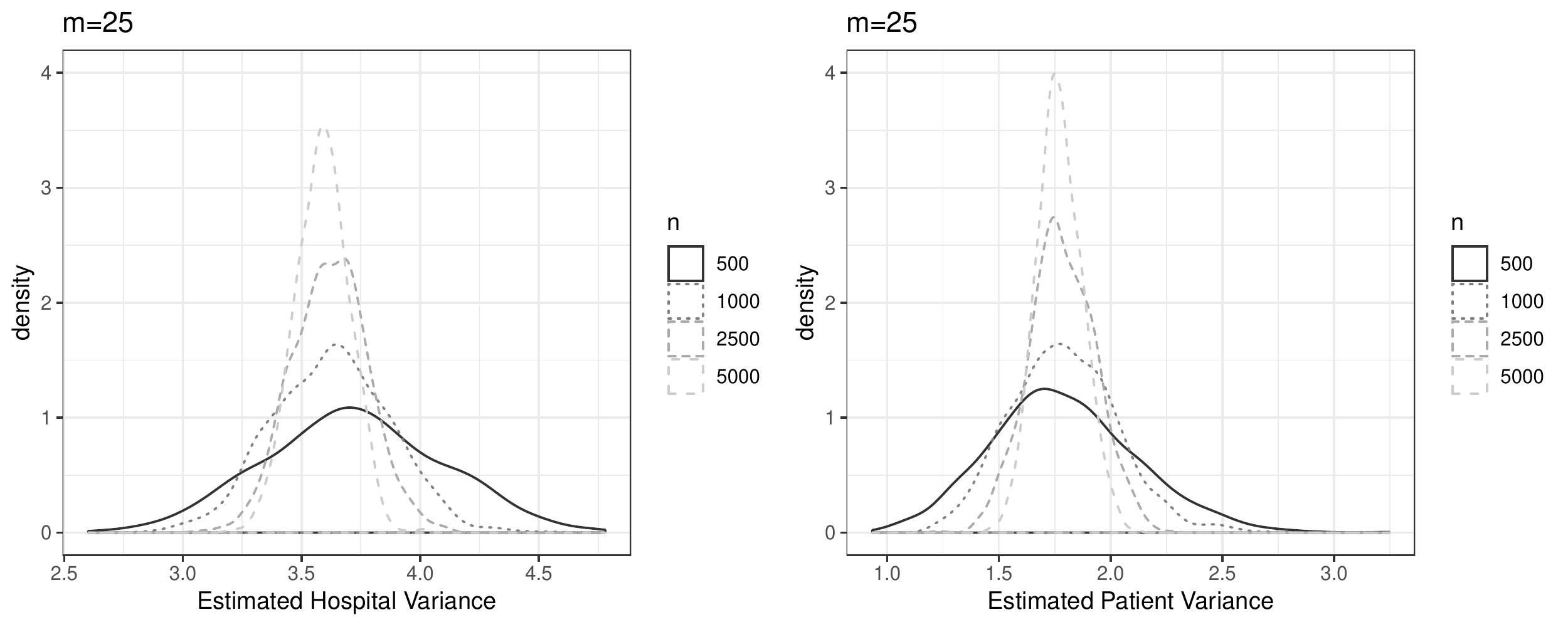}
\caption{Density plots for the simulated sampling distributions of the case-mix and between-hospital variance components with a continuous outcome, based on 1000 replications. $n$ and $m$ are the total number of patients and total number of hospitals.}
\label{Plot2_con}
\end{figure}

\begin{figure}[!h]
\centering
\includegraphics[width=100mm]{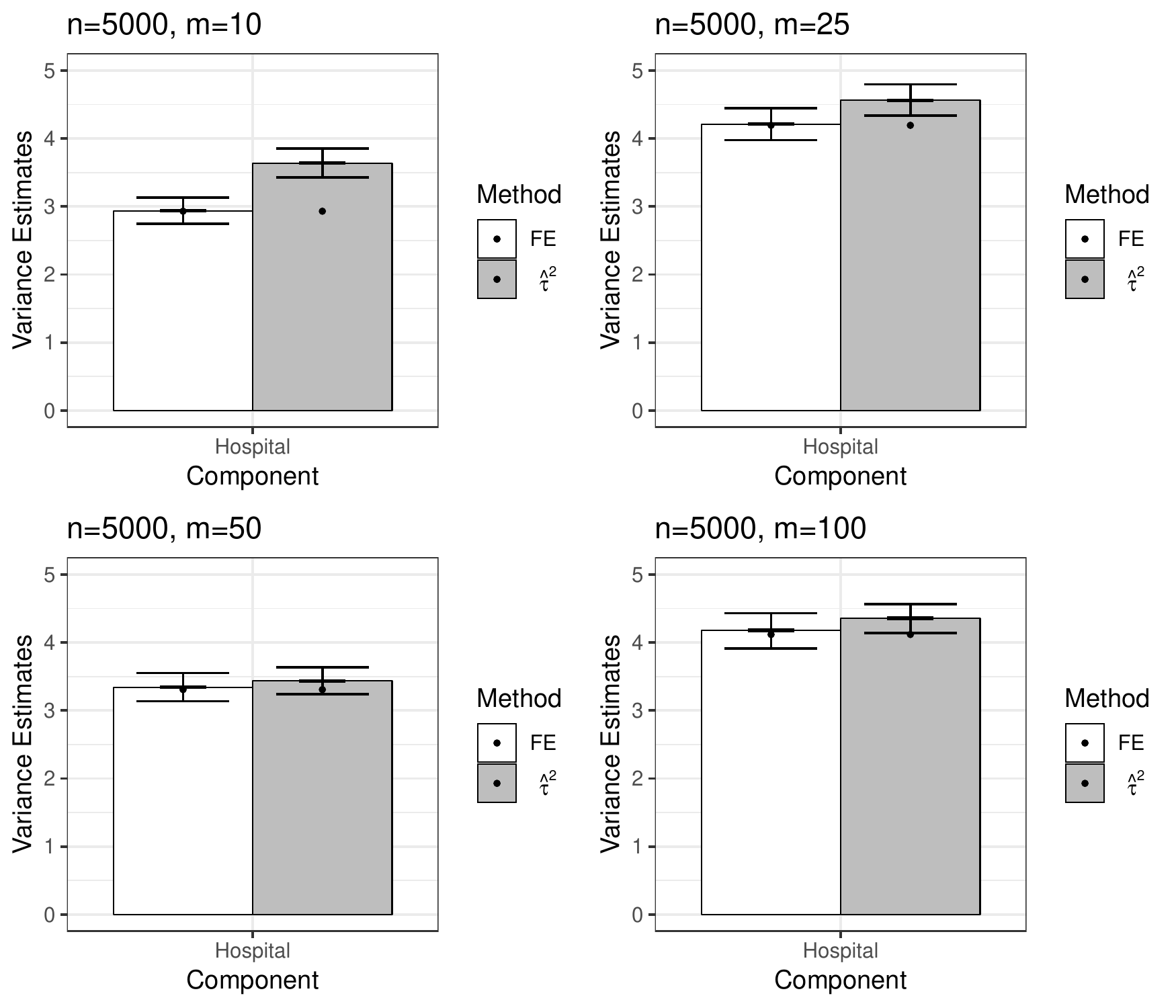}
\caption{Simulated sampling distribution means for the between-hospital variance components, using the proposed estimator, and random effect variance $\tau^2$, based on 1000 replications. FE stands for fixed-effect model. $n$ and $m$ are the total number of patients and total number of hospitals.}
\label{Plot3_con}
\end{figure}

Figure \ref{Plot1_con} shows the simulated sampling distribution means for the three variance components under both fixed and random effect models for the continuous outcomes under different combinations of $n$ and $m$, based on 1000 replications. Also shown are the 95\% quantile interval for the sampling distribution, and the 95\% confidence interval for the mean of the sampling distribution, reflecting  the magnitude of the Monte Carlo error. The black dots indicate the true variances. The estimators capture the true value reasonably well in all cases, though with the random effect model giving slightly smaller estimates for the hospital variance component. Also of note is that the uncertainty in the variance component estimates is mainly driven by the overall number of patients $n$ (distributed across the $m$ hospitals) rather than the number of hospitals $m$. This is also illustrated in Figure \ref{Plot2_con} which shows the convergence of the sampling distributions for the hospital and patient variance components as a function of $n$ with fixed $m$. Finally, as suggested in Section \ref{section:intraclass}, in Figure \ref{Plot3_con} we demonstrate that the random effect variance $\hat\tau^2$ estimates a different quantity than the causal between-hospital variance in the decomposition; the difference is marked especially with small number of hospitals.

\subsubsection{Binary Outcomes}

\begin{figure}[!h]
\centering
\includegraphics[width=130mm]{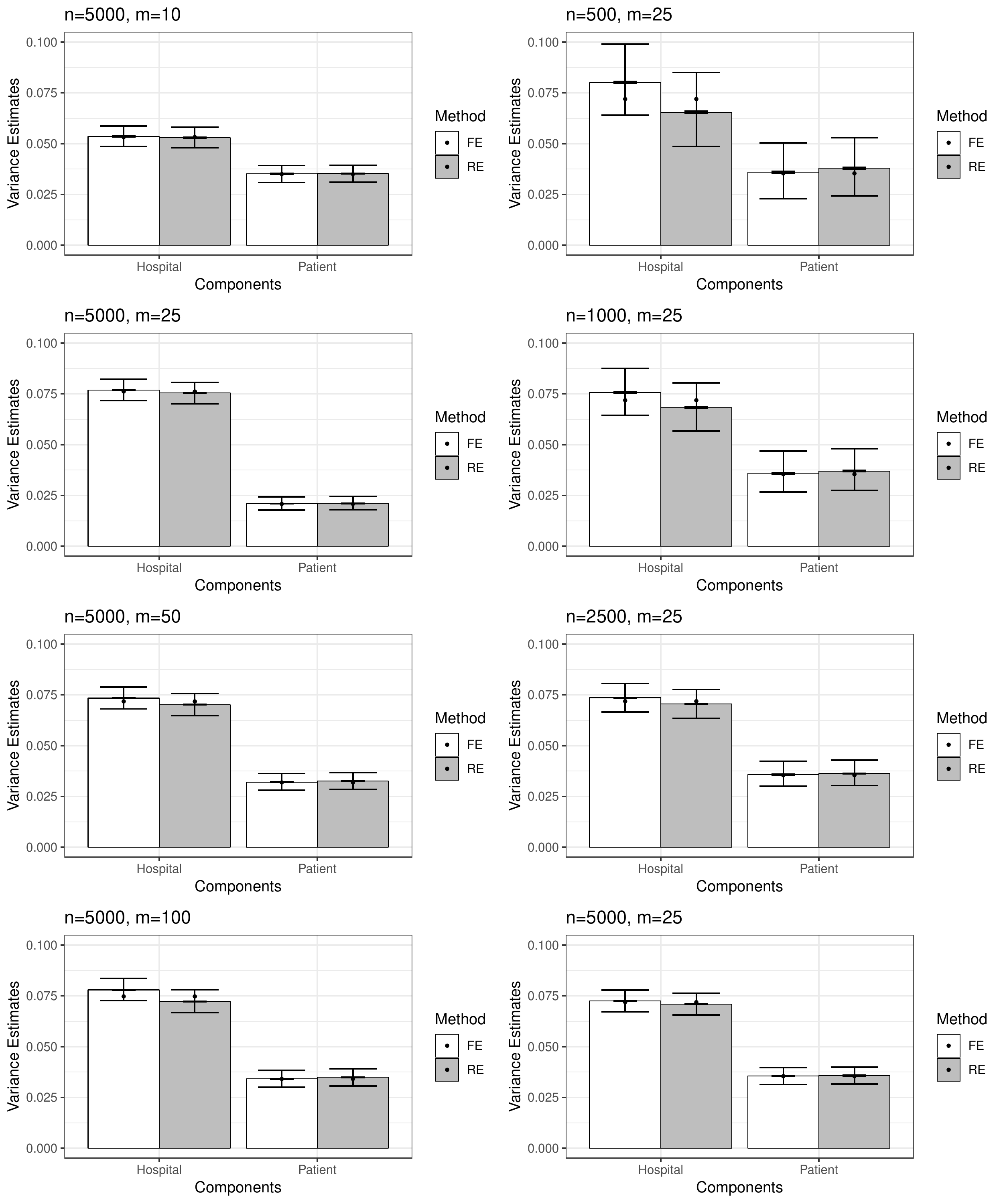}
\caption{Simulated sampling distribution means for the case-mix and between-hospital variance components under both fixed and random effect models for the binary outcomes under different combinations of $n$ (total number of patients) and $m$ (total number of hospitals), based on 1000 replications. Also shown are the 95\% quantile interval for the sampling distribution, and the 95\% confidence interval for the mean. The black dots indicate the true variances. FE and RE stand for fixed-effect and random-effect models.}
\label{Plot1_binary}
\end{figure}

\begin{figure}[!h]
\centering
\includegraphics[width=130mm]{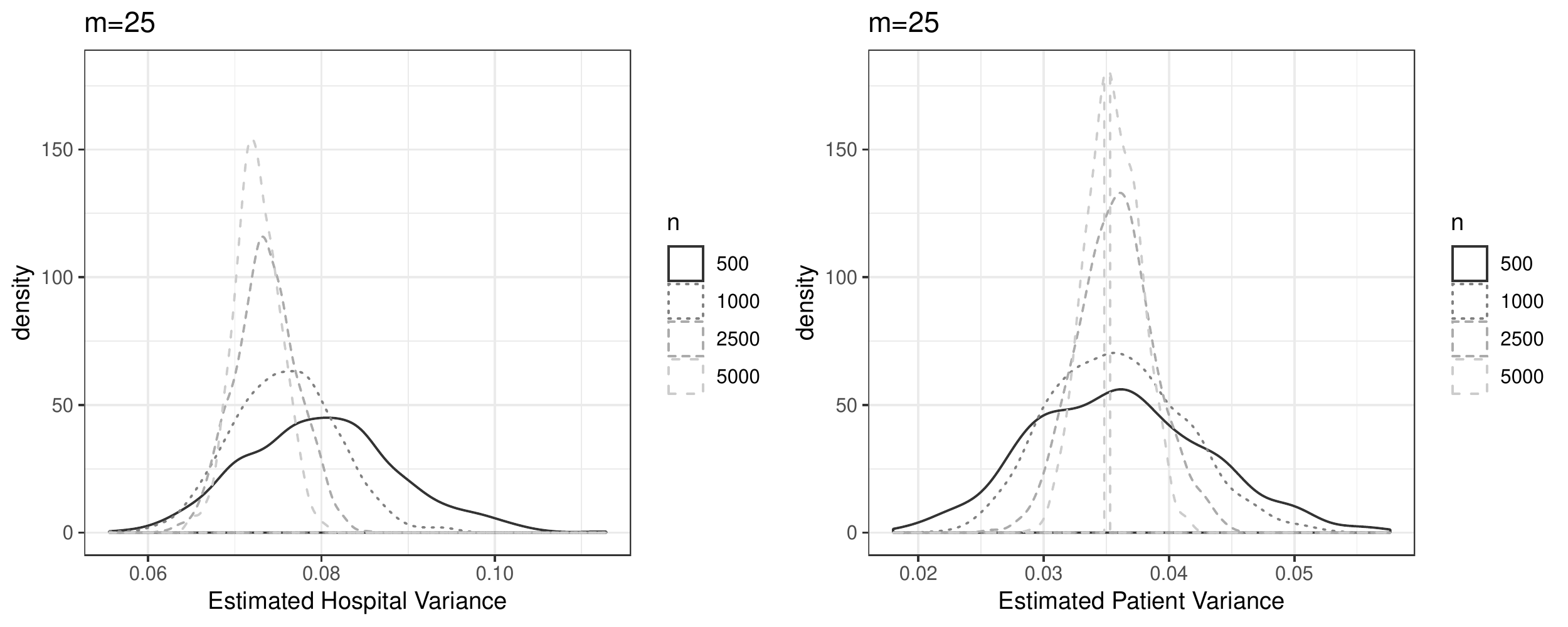}
\caption{Density plots for the simulated sampling distributions of the case-mix and between-hospital variance components with a binary outcome, based on 1000 replications. $n$ and $m$ are the total number of patients and total number of hospitals.}
\label{Plot2_binary}
\end{figure}

Figure \ref{Plot1_binary} shows the simulated sampling distribution means for two of the three variance components under both fixed and random effect models for the binary outcomes under different combinations of $n$ and $m$, based on 1000 replications. The results are mostly similar to the continuous case, with the random effect model giving slightly smaller estimates for the hospital variance component compared to the fixed effect model. However, some small sample bias in the hospital variance component (fixed effect model overestimating and random effect model underestimating) is visible with small number of patients per hospital (top-right panel, on average $500/25=20$ patients per hospital), which disappears with increasing number of patients. The convergence of the sampling distributions with increasing $n$ is illustrated in the density plots of Figure \ref{Plot2_binary}.

\section{Illustration in real data}\label{section:illustration}

We demonstrate the use of the proposed three-way causal decomposition in the context of QIs for surgical care of kidney cancer in the province of Ontario. We investigate four processes/outcomes that have previously been proposed as disease-specific QIs using the consensus/Delphi method by a panel of experts \cite{wood:2013}. These are the proportion of partial (versus radical) nephrectomies among stage T1a nephrectomy patients (Partial), the same proportion restricted to the subpopulation of patients with chronic kidney disease (CKD) or its risk factors diabetes or hypertension (Partial-CKD), minimally invasive (versus open) surgery among T1-T2 radical nephrectomy patients (MIS), and readmission within 30 days of the surgery for T1-T4 radical nephrectomy patients (Readmission). The Partial, Partial-CKD and MIS indicators are process type, with higher proportion reflecting quality of care through kidney function conserving and less invasive treatment for early stage patients that can benefit from this. The Readmission indicator is outcome type, with lower proportion reflecting quality of care through reduced complications. Our objective in the illustration was to evaluate the processes/outcomes using the proposed variance decomposition approach, to quantify how much of the variation in these is between-hospital (reflecting performance and potentially intervenable), and how much is driven by the patient case-mix. The data were obtained from cross-linkage of Cancer registry, pathology reports, Hospital Discharge Abstract (DAD), Ontario Health Insurance Plan (OHIP) and other databases housed at the Institute for Clinical Evaluative Sciences (ICES). We included patients who had complete data on the case-mix factors age, sex, income quintile, Charlson comorbidity score, ACG comorbidity score, presence of CKD risk factors (for indicators other than Partial-CKD), days from diagnosis to treatment, year of diagnosis, tumor size, and T stage. Table \ref{cov} shows the number $n$ of patients identified and hospitals $m$ evaluated for each of the four indicators in the study period from 1995-2014.

To these data, using the R package \texttt{VGAM} \cite{vgam}, we fitted a multinomial logistic assignment models where, to deal with empty covariate categories for small volume hospitals, we estimated only intercept terms for hospitals that had treated less than 35 (40 for MIS) patients over the study period. This is justifiable as our estimators use the assignment model only for weighting the hospitals by their patient volumes, whereas the case-mix adjustment is achieved through the outcome model. For the processes/outcomes, we fitted mixed effects logistic models using the \texttt{glmer} function of R package \texttt{lme4} \cite{lme4}, adjusting for the aforementioned covariates. The covariate effects (log-odds ratios) and their z-scores are reported in Table \ref{cov}, along with the likelihood ratio test $\chi^2$ values for the random intercept variance parameter in the models (one degree of freedom). In all cases the likelihood ratio test indicated statistically significant between hospital variation, but this was much smaller for Readmission. Larger tumor size was a very strong predictor of radical nephrectomy, while both partial nephrectomies and minimally invasive surgeries had become significantly more common over calendar time. The comorbidity scores were strongest predictors of readmission.

The empirical proportions of the four indicators and the corresponding marginal variances were 44.7\% and 0.447*(1-0.447)=0.247 (Partial), 43.8\% and 0.246 (Partial-CKD), 63.7\% and 0.231 (MIS), and 7.6\% and 0.07 (Readmission). Table \ref{real_data} shows the estimated three variance components using the proposed decomposition, along with the 95\% credible intervals using the approximate Bayesian method described in Section \ref{section:estimation}. We also show the proportional variances along with their credible intervals, because these are more informative for comparing indicators with very different marginal variance. We note that the estimated components sum up to the empirical marginal variance given by the binomial calculation. The Partial indicator shows some between-hospital variation, but this is small compared to the case-mix variance. This does not change much when this indicator is restricted to the CKD risk factors subpopulation, because tumor size is the major driver of the large case-mix variance. These results suggest that even though the indicator was restricted to T4a disease, the treatment choice seems to be often determined by disease progression. However, because partial proportion has seen significant increase over calendar time, the case-mix variance will be smaller when such indicator is applied to contemporary data. In contrast, the MIS indicator shows between-hospital variance that is almost two times the case-mix variance, implying lot of variation in care for similar patients. This may be due to specialization or facilities available, which are intervenable. Because the calendar year was the biggest driver of the case-mix variance for MIS, the case-mix variance will be even smaller when indicator is applied to contemporary data. Finally, the Readmission indicator shows very little between-hospital variation, suggesting it is unlikely to be useful in capturing performance related variation.

\begin{table}
\begin{center}
\begin{tabular}{lcccc}
\multicolumn{5}{c}{Process/outcome} \\ \hline
\multirow{4}{*}{Covariate} & Partial          & Partial-CKD       & MIS               & Readmission     \\ 
& $n=4738$ & $n=3028$  & $n=4100$ & $n=7362$ \\
& $m=112$  & $m=110$    & $m=72$    & $m=115$  \\ 
& $\chi^2=377$  & $\chi^2=216$    & $\chi^2=825$    & $\chi^2=16$  \\ \hline

Male sex                                                                                       & 0.21 (2.81)                 & 0.22 (2.39)                       & 0.05 (0.67)       & -0.02 (-0.18)               \\
Age 			                                                                          & -0.01 (-4.11)               & -0.01 (-2.67)                     & -0.01 (-2.06)     & 0.01 (2.74)               \\
Income quintile                                                                    & 0.03 (1.14)                 & 0.03 (1.07) & 0.01 (0.47)       & \textless\textbar0.01\textbar (-0.05) \\
Rural vs urban                                                                     & 0.18 (1.46)                 & 0.20 (1.37)                       & -0.14 (-1.17)     & -0.18 (-1.33)               \\
Charlson score                                                                     & -0.14 (-4.93)               & -0.16 (-5.02)                     & -0.11 (-4.37)     & 0.12 (6.56)                 \\
ACG score                                                                             & -0.01 (-1.20) & \textless\textbar0.01\textbar (-0.19)            & 0.01 (1.61)       & 0.02 (4.64)                 \\
log(dx to tx days+1)                                                          & 0.02 (1.02)                 & 0.03 (1.41)                       & 0.02 (0.65)       & 0.09 (3.64)                 \\
Year of dx                                                                            & 0.19 (21.87)                & 0.18 (17.13)                      & 0.20 (15.28)      & -0.01 (-1.43)               \\
Tumor size (cm)                                                                            & -0.82 (-17.25)              & -0.75 (-12.87)                    & -0.16 (-7.21)     & 0.02 (1.44)                 \\
CKD risk factors                                                                      & -0.02 (-0.26)             & -                                  & -0.03 (-0.36)     & 0.15 (1.31)                 \\
T2 vs T1 stage                                                                    &  -                           & -                                  & -0.31 (-2.16)      & -0.03 (-0.21)               \\ 
T3 or 4 vs T1 stage                                                          &  -                          & -                                  & -                  & 0.19 (-1.60)
\end{tabular}
\caption{A summary table of the effects of case-mix factors for the four different QIs. The numbers are log-odds ratios (z-scores), while $m$ and $n$ represent the number of hospitals and patients. dx stands for diagnosis and tx for treatment (surgery).}
\label{cov}
\end{center}
\end{table}

\begin{table}
\begin{center}
\begin{tabular}{ccccc}
\multicolumn{5}{c}{Source of variation} \\ \hline
Process/outcome                            &     & Patient case-mix   & Between-hospital & Residual             \\ \hline
\multirow{4}{*}{Partial}          & Variance                 & $0.067$            & $0.017$                   & $0.164$              \\
                                             & $95\%$ CI & $(0.062, 0.072)$   & $(0.014, 0.020)$          & $(0.158, 0.169)$     \\
                                             & Proportion      & $26.9\%$           & $6.9\%$                   & $66.2\%$             \\
                                             & $95\%$ CI & $(24.9\%, 29.0\%)$ & $(5.5\%, 8.1\%)$          & $(64.0\%, 68.5\%)$   \\ \cline{2-5} 
\multirow{4}{*}{Partial-CKD} & Variance                 & $0.058$            & $0.016$                   & $0.173$              \\
                                             & $95\%$ CI & $(0.052, 0.065)$   & $(0.012, 0.020)$          & $(0.165, 0.179)$     \\
                                             & Proportion      & $23.7\%$           & $6.3\%$                   & $70.0\%$             \\
                                             & $95\%$ CI & $(21.2\%, 26.3\%)$ & $(4.7\%, 8.1\%)$          & $(67.1\%, 72.7\%)$   \\ \cline{2-5} 
\multirow{4}{*}{MIS}                         & Variance                 & $0.023$            & $0.044$                   & $0.164$              \\
                                             & $95\%$ CI & $(0.019, 0.027)$   & $(0.039, 0.050)$          & $(0.158, 0.170)$     \\
                                             & Proportion      & $9.9\%$           & $19.0\%$                  & $71.1\%$             \\
                                             & $95\%$ CI & $(8.1\%, 11.5\%)$  & $(16.8\%, 21.6\%)$        & $(68.4\%, 73.7\%)$   \\ \cline{2-5} 
\multirow{4}{*}{Readmission}      & Variance                 & $0.0020$           & $0.0001$                  & $0.0679$             \\
                                             & $95\%$ CI & $(0.0013, 0.0027)$ & $(<0.0001, 0.0004)$       & $(0.0670, 0.0685)$   \\
                                             & Proportion      & $2.81\%$           & $0.16\%$                  & $97.03\%$            \\
                                             & $95\%$ CI & $(1.92\%, 3.92\%)$ & $(<0.01\%, 0.55\%)$       & $(95.74\%, 97.94\%)$
\end{tabular}
\caption{Three-way variance decomposition for the observed variance in the four different QIs. The numbers are absolute and proportional variances (95\% CI = 95\% credible interval).}
\label{real_data}
\end{center}
\end{table}

\section{Discussion}\label{section:discussion}

In this paper, we aimed to give a causal interpretation to variance decompositions used to quantify variations in hospital performance, and propose estimators that can accommodate different link functions, as well as fixed and random effect models for the hospital effects. The estimation of the hospital variance component worked reasonably well with the number of hospitals as small as 10. While the random effect models could introduce some bias to estimation of the variance components when the hospital-level effects and the patient case-mix are dependent \cite{Joseph2014}, they are helpful for smoothing purposes when a large number of small volume hospitals are present. Fixed effect models can accommodate interactions between case-mix factors and hospital effects, but this leads to a large number of parameters to be estimated \cite{Varewyck:2016}.

Addressing the variance decompositions in an explicit causal modeling framework helps to answer questions regarding the part of the variation in care received that is explained by different practices for similar patients. Furthermore, quantifying explained confounding due to observed case-mix factors into its own variance component helps to assess the usefulness of a given process or outcome in constructing a quality indicator. Also, we note that, similar to random effect meta-analysis, the estimation of the between-hospital variance component is weighted roughly in proportion to the hospital-specific patient volumes, meaning that the unstable estimates of small hospitals do not have undue influence on the estimation of this variance component.

Several extensions are possible. We split the total variance into that explained by hospital performance, case-mix, and unexplained residual variance. However, we did not separate between case-mix factors that capture disease progression and thus directly (and justifiably) influence the treatment decisions, and those capturing disparities in care, such as sociodemographic and socioeconomic factors. While the latter can be confounders under the causal model, any variation in care received that is explained by such factors may be of interest in itself, and thus could be split into a separate variance component. Of particular interest might be interactions between hospital effects and sociodemographic/economic factors, which would imply that disparities in care occur within hospitals.

In this paper we considered only a single layer of exposure hierarchy, such as hospitals. We are currently working on extending the causal variance decompositions to multiple layers of a hierarchy, such as surgeons within hospitals or referral networks within administrative subregions. This will require introduction of multiple levels of exposure, using notation similar to causal mediation analysis models \cite{VanderWeele:2009b}, where the causal pathway leads, for example, from hospital to surgeon to outcome. This will enable us to further decompose the hospital effects to those due to institution level factors and those resulting from practitioner level differences. However, since the models with nested exposures are made identifiable through random effects, this will require further consideration of the role of random effects in causal models.

\subsection*{Funding}

This work was supported by a Discovery Grant from the Natural Sciences and Engineering Research Council of Canada (to OS), a Catalyst Grant in Health Services and Economics Research from the Canadian Institutes of Health Research (to AF, KAL and OS) and the Ontario Institute for Cancer Research through funding provided by the Government
of Ontario (to BC).

\bibliographystyle{unsrt}

\begin{thebibliography}{10}
\providecommand{\url}[1]{\texttt{#1}}
\providecommand{\urlprefix}{URL }
\expandafter\ifx\csname urlstyle\endcsname\relax
  \providecommand{\doi}[1]{DOI:\discretionary{}{}{}#1}\else
  \providecommand{\doi}{DOI:\discretionary{}{}{}\begingroup
  \urlstyle{rm}\Url}\fi
\providecommand{\eprint}[2][]{\url{#2}}

\bibitem{donabedian:1988}
Donabedian A.
\newblock {{T}he quality of care. {H}ow can it be assessed?}
\newblock \emph{JAMA} 1988; 260(12): 1743--1748.

\bibitem{wood:2013}
Wood L, Bjarnason GA, Black PC et~al.
\newblock {{U}sing the {D}elphi technique to improve clinical outcomes through
  the development of quality indicators in renal cell carcinoma}.
\newblock \emph{J Oncol Pract} 2013; 9(5): e262--267.

\bibitem{khuri:1998}
Khuri SF, Daley J, Henderson W et~al.
\newblock {{T}he {D}epartment of {V}eterans {A}ffairs' {N}{S}{Q}{I}{P}: the
  first national, validated, outcome-based, risk-adjusted, and peer-controlled
  program for the measurement and enhancement of the quality of surgical care.
  {N}ational {V}{A} {S}urgical {Q}uality {I}mprovement {P}rogram}.
\newblock \emph{Ann Surg} 1998; 228: 491--507.

\bibitem{varewyck:2014}
Varewyck M, Goetghebeur E, Eriksson M et~al.
\newblock On shrinkage and model extrapolation in the evaluation of clinical
  center performance.
\newblock \emph{Biostatistics} 2014; 15(4): 651--664.

\bibitem{Houwelingen1999EmpiricalBM}
van Houwelingen HC and Brand R.
\newblock Empirical bayes methods for monitoring health care quality.
\newblock In \emph{Bulletin of the ISI, ISI 99, Book 1}. pp. 75--78.

\bibitem{dishoeck:2011}
van Dishoeck AM, Lingsma HF, Mackenbach JP et~al.
\newblock Random variation and rankability of hospitals using outcome
  indicators.
\newblock \emph{BMJ Quality and Safety} 2011; 20(10): 869--874.

\bibitem{Henneman:2014}
Henneman D, van Bommel ACM, Snijders A et~al.
\newblock Ranking and rankability of hospital postoperative mortality rates in
  colorectal cancer surgery.
\newblock \emph{Annals of Surgery} 2014; 259(5): 844--849.

\bibitem{racz:2010}
Racz MJ and Sedransk J.
\newblock {Bayesian and frequentist methods for provider profiling using
  risk-adjusted assessments of medical outcomes}.
\newblock \emph{Journal of the American Statistical Association} 2010; 105:
  48--58.

\bibitem{dersimonian:1987}
DerSimonian R and Laird N.
\newblock {{M}eta-analysis in clinical trials}.
\newblock \emph{Control Clin Trials} 1986; 7(3): 177--188.

\bibitem{Ridout:1999}
Ridout MS, Dem{\'e}trio CGB and Firth D.
\newblock Estimating intraclass correlation for binary data.
\newblock \emph{Biometrics} 1999; 55(1): 137--148.

\bibitem{Merlo:2006}
Merlo J, Chaix B, Ohlsson H et~al.
\newblock A brief conceptual tutorial of multilevel analysis in social
  epidemiology: using measures of clustering in multilevel logistic regression
  to investigate contextual phenomena.
\newblock \emph{Journal of Epidemiology and Community Health} 2006; 60(4):
  290--297.

\bibitem{Andrew2009}
Thomson A, Hayes R and Cousens S.
\newblock Measures of between-cluster variability in cluster randomized trials
  with binary outcomes.
\newblock \emph{Statistics in Medicine} 2009; 28(12): 1739--1751.

\bibitem{Yelland2011}
Yelland LN, Salter AB, Ryan P et~al.
\newblock Adjusted intraclass correlation coefficients for binary data: methods
  and estimates from a cluster-randomized trial in primary care.
\newblock \emph{Clinical Trials} 2011; 8(1): 48--58.

\bibitem{Wu:2012}
Wu S, Crespi CM and Wong WK.
\newblock Comparison of methods for estimating the intraclass correlation
  coefficient for binary responses in cancer prevention cluster randomized
  trials.
\newblock \emph{Contemporary Clinical Trials} 2012; 33(5): 869--880.

\bibitem{Bo:2017}
Chen B and Benedetti A.
\newblock Quantifying heterogeneity in individual participant data
  meta-analysis with binary outcomes.
\newblock \emph{Systematic Reviews} 2017; 6(1): 243.

\bibitem{Thomas2013}
Debray TPA, Moons KGM, Abo-Zaid GMA et~al.
\newblock Individual participant data meta-analysis for a binary outcome:
  One-stage or two-stage?
\newblock \emph{PLoS One} 2013; 8(4): e60650.

\bibitem{demaris:2002}
DeMaris A.
\newblock Explained variance in logistic regression: A monte carlo study of
  proposed measures.
\newblock \emph{Sociological Methods and Research} 2002; 21(1): 27--74.

\bibitem{Goldstein:2002}
Goldstein H, Browne W and Rasbash J.
\newblock Partitioning variation in multilevel models.
\newblock \emph{Understanding Statistics} 2002; 1(4): 223--231.

\bibitem{austin2017}
Austin PC, Wagner P and Merlo J.
\newblock The median hazard ratio: a useful measure of variance and general
  contextual effects in multilevel survival analysis.
\newblock \emph{Statistics in Medicine} 2017; 36(6): 928--938.

\bibitem{bowsher:2012}
Bowsher CG and Swain PS.
\newblock Identifying sources of variation and the flow of information in
  biochemical networks.
\newblock \emph{Proceedings of the National Academy of Sciences} 2012; 109(20):
  E1320--E1328.

\bibitem{rosenbaum:1983}
Rubin DB and Rosenbaum PR.
\newblock The central role of the propensity score in observational studies for
  causal effects.
\newblock \emph{Biometrika} 1983; 70(1): 41--55.

\bibitem{Miguel:2006}
Hern{\'a}n MA and Robins JM.
\newblock Estimating causal effects from epidemiological data.
\newblock \emph{Journal of Epidemiology and Community Health} 2006; 60:
  578--586.

\bibitem{moreno:2017}
Moreno-Betancur M, Koplin JJ, Ponsonby AL et~al.
\newblock Measuring the impact of differences in risk factor distributions on
  cross-population differences in disease occurrence: a causal approach.
\newblock \emph{International Journal of Epidemiology} 2018; 47(1): 217--225.

\bibitem{VanderWeele:2014}
VanderWeele T, Vansteelandt S and Robins J.
\newblock {{E}ffect decomposition in the presence of an exposure-induced
  mediator-outcome confounder}.
\newblock \emph{Epidemiology} 2014; 25(2): 300--306.

\bibitem{Varewyck:2016}
Varewyck M, Vansteelandt S, Eriksson M et~al.
\newblock On the practice of ignoring center-patient interactions in evaluating
  hospital performance.
\newblock \emph{Stat Med} 2016; 35: 227--238.

\bibitem{lme4}
Bates D, M{\"a}chler M, Bolker B et~al.
\newblock Fitting linear mixed-effects models using {lme4}.
\newblock \emph{Journal of Statistical Software} 2015; 67(1): 1--48.
\newblock \doi{10.18637/jss.v067.i01}.

\bibitem{vgam}
Yee TW, Stoklosa J and Huggins RM.
\newblock The {VGAM} package for capture-recapture data using the conditional
  likelihood.
\newblock \emph{Journal of Statistical Software} 2015; 65(5): 1--33.
\newblock \urlprefix\url{http://www.jstatsoft.org/v65/i05/}.

\bibitem{Joseph2014}
Dieleman JL and Templin T.
\newblock Random-effects, fixed effects and the within-between specification
  for clustered data in observational health studies: A simulation study.
\newblock \emph{PLoS One} 2014; 9(10): e110257.

\bibitem{VanderWeele:2009b}
VanderWeele T and Vansteelandt S.
\newblock Conceptual issues concerning mediation, interventions and
  composition.
\newblock \emph{Stat Interface} 2009; 2: 457--468.

\end{thebibliography}

\end{document}